\documentclass[prl,twocolumn,superscriptaddress,floatfix,nopacs]{revtex4}
\usepackage{graphicx,amsfonts,amssymb,amsmath,hyperref,enumerate,color}
\usepackage{bbold}

 \definecolor{jens}{rgb}{0.2,0.8,0.7}
   
 \definecolor{matteo}{rgb}{0.4,0.0,0.8}
   
\definecolor{roman}{rgb}{1,0,0}
   
\definecolor{augustine}{rgb}{0,0,1}
      
\newif\ifhyper
\hypertrue
\ifhyper
\hypersetup{
   citecolor = {green},
   colorlinks = {true}, 
   urlcolor = {blue} 
}
\fi

\newcommand{\beq}{\begin{equation}}
\newcommand{\eeq}{\end{equation}}
\newcommand{\beqa}{\begin{eqnarray}}
\newcommand{\eeqa}{\end{eqnarray}}
\newcommand{\ket} [1] {\vert #1 \rangle}

\def\ket#1{\vert#1\rangle}

\def\Longarrow{\protect\@lra}
\def\@lra{\relbar\joinrel\relbar\joinrel\relbar\joinrel%
          \relbar\joinrel\rightarrow}

\usepackage{times}
\begin{document}

\title{A tensor network annealing algorithm for two-dimensional thermal states}

\author{A.\
Kshetrimayum}
\affiliation{Dahlem Center for Complex Quantum Systems, Physics Department, Freie Universit\"{a}t Berlin, 14195 Berlin, Germany}
\affiliation{Institute of Physics, Johannes Gutenberg University, 55099 Mainz, Germany}

\author{M.\
Rizzi}
\affiliation{Institute of Physics, Johannes Gutenberg University, 55099 Mainz, Germany}

\author{J.\
Eisert}
\affiliation{Dahlem Center for Complex Quantum Systems, Physics Department, Freie Universit\"{a}t Berlin, 14195 Berlin, Germany}
\affiliation{Department of Mathematics and Computer Science, Freie Universit\"{a}t Berlin, 14195 Berlin, Germany}

\author{R.\
Or\'us}
\affiliation{Institute of Physics, Johannes Gutenberg University, 55099 Mainz, Germany}
\affiliation{Donostia International Physics Center, Paseo Manuel de Lardizabal 4, E-20018 San Sebasti\'an, Spain}
\affiliation{Ikerbasque Foundation for Science, Maria Diaz de Haro 3, E-48013 Bilbao, Spain}

\begin{abstract}
Tensor network methods have become a powerful class of tools to capture strongly correlated matter, but methods to capture the experimentally ubiquitous family of models 
at finite temperature beyond one spatial dimension are largely lacking. We introduce a tensor network algorithm able to simulate thermal states of two-dimensional quantum lattice systems in the thermodynamic limit. The  method develops instances of projected entangled pair states and projected entangled pair operators for this purpose. It is the key feature of this algorithm to resemble the cooling down of the system from an infinite temperature state until it reaches the desired finite-temperature regime. As a benchmark we study the finite-temperature phase transition of the Ising model on an infinite square lattice, for which we obtain remarkable agreement with the exact solution. We then turn to study the finite-temperature Bose-Hubbard model in the limits of two (hard-core) and three bosonic modes per site. Our technique can be used to support the experimental study of actual effectively two-dimensional materials in the laboratory, as well as to benchmark optical lattice quantum simulators with ultra-cold atoms.
\end{abstract}

\date{\today}

\maketitle

\emph{Tensor network (TN)} algorithms have become a powerful tool in the study of quantum many-body systems
\cite{Orus-AnnPhys-2014,VerstraeteBig,TensorNetworkReview,AreaReview}. 
They build upon and 
further develop the so-called \emph{density matrix renormalisation group (DMRG)} approach \cite{MPSRev,DMRGWhite92},  
that is able to 
simulate ground states of one-dimensional strongly correlated systems essentially to machine precision. Once the 
structure of DMRG was understood as being a tensor network approach \cite{Rommer}, further method development followed.
This prominently included 
methods to capture time-dependent problems 
\cite{White2004,Feiguin2005,Daley}, as well 
as a machinery to describe open dissipative systems \cite{PhysRevLett.93.207204,PositiveMPO,Kshetrimayum,PhysRevA.87.033606,DaleyOpenReview}
and thermal states in one spatial dimension \cite{MinimalEntanglement,PhysRevB.94.115157,Weichselbaum}. 

The study of two dimensional strongly correlated systems with TN methods, however, comes along with serious numerical effort and 
conceptual challenges \cite{Contraction,PhysRevA.95.060102}.
\emph{Projected entangled pair states (PEPS)} allow to grasp pure ground states of two-dimensional models \cite{PEPSOld,simpleupdatejordan}. For thermal states, though, the numerical challenge is even harder, where only few methods have been proposed for simple spin systems~\cite{piotr2012,piotr2015,piotr2016}  in 
sharp contrast to ground state calculations and much remains to be explored. This is even more of a serious omission since two-dimensional quantum systems at finite temperature are ubiquitous in a number of context. This prominently includes effectively two-dimensional \emph{quantum materials} in real laboratory 
conditions as well as systems of ultra-cold atoms in optical lattices in instances of \emph{quantum
simulations} \cite{CiracZollerSimulation,BlochSimulation} in the quantum technologies \cite{Roadmap}. 

In this work, we 
innovate an efficient tensor network algorithm for capturing thermal states of quantum lattice systems in two spatial dimensions and in the thermodynamic limit. Our approach 
significantly further  develops a core idea of
Ref.\ \cite{Kshetrimayum} in that it uses the vectorization of a PEPS together with simple update and \emph{corner transfer matrix (CTM)} 
techniques \cite{ctmnishino1996,ctmnishino1997,ctmroman2009,ctmroman2012}. The scheme starts from an infinite-temperature state and simulates the annealing towards lower temperatures until reaching the desired regime. Our method is particularly efficient and practical in realistic situations. Compared to previous attempts to simulate 2D thermal states with TNs, it can go well beyond
Ising-type simulations of locally-purified thermal PEPS \cite{PhysRevB.92.035152}, and is far more efficient than 2D TN algorithms based on cluster updates and self-consistent environment calculations \cite{shiju}.
To exemplify this point, equipped with this powerful tool, we turn to
simulating
 thermal Ising and Bose-Hubbard models on an infinite square lattice. While the Ising model is an excellent benchmark \cite{onsager}, the study of the 
Bose-Hubbard model allows to compare with realistic lab settings in optical lattice experiments performing quantum simulations 
\cite{BlochSimulation,CiracZollerSimulation,Roadmap} of strongly correlated matter.

\begin{figure}
 	\includegraphics[width=0.5\textwidth]{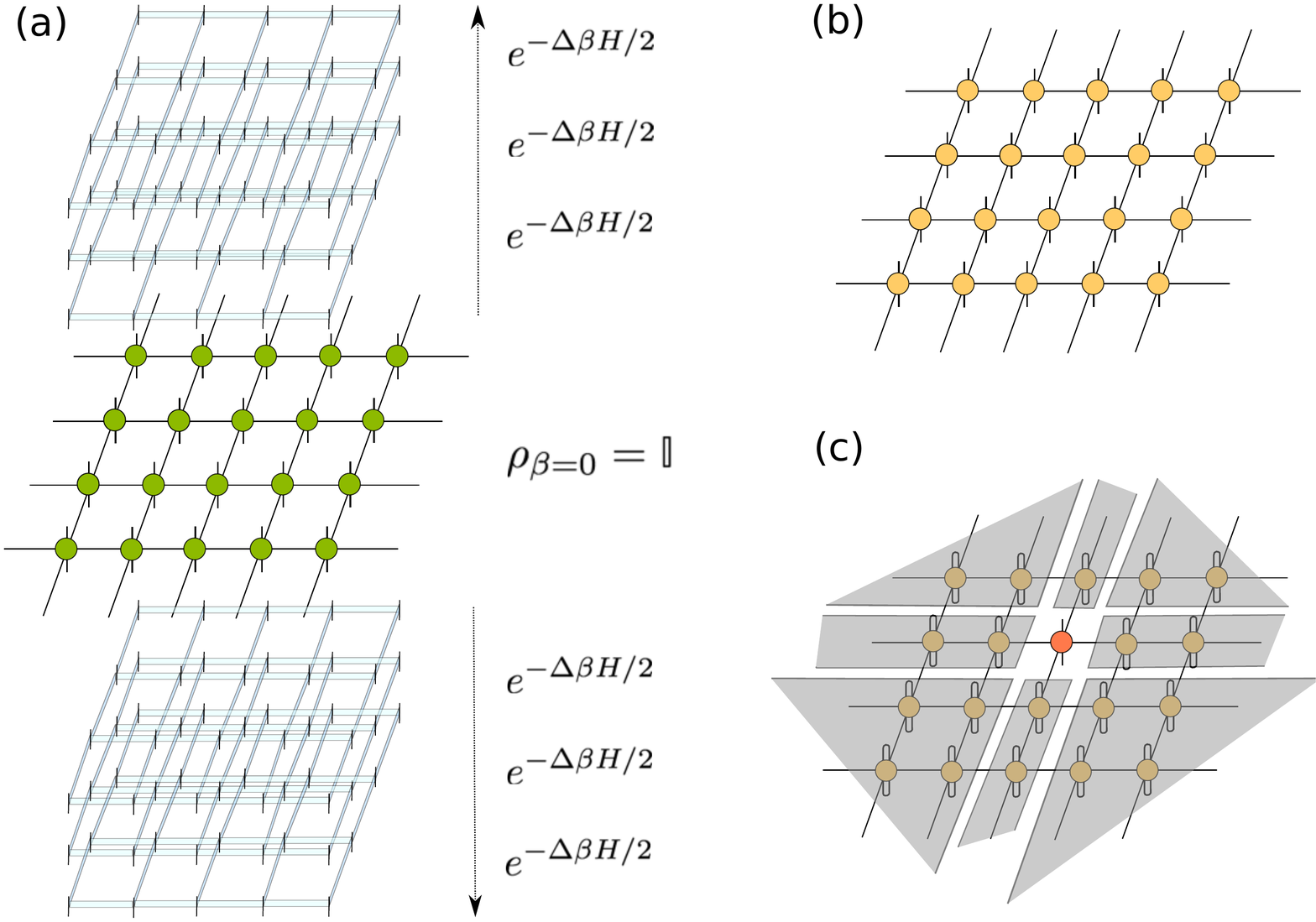}
 	\caption{(a) ``Slicing" of an unnormalized thermal state $\rho = e^{- \beta H}$ in $m$ steps with inverse temperature change $\Delta \beta$, as explained in the main text, and for an infinite 2D square lattice. The evolution operator $e^{- \Delta \beta H}$ is applied to both bra and ket indexes simultaneously, starting from the infinite-temperature mixed state ${\mathbb I}$. After each ``slice", the outcome can be approximated by a PEPO as in (b). In (c) we show the contraction needed to compute a one-site reduced density matrix, needed to evaluate local expectation values on that site. This contraction is done using CTM methods. The structure of the tensors leading to the CTMs is shown in the shaded region.}
 	\label{pepo1}
\end{figure}

\emph{Methods.}  Our approach is a ``cooling down" or ``annealing" technique that we describe in what follows. A thermal state in the canonical ensemble and for Hamiltonian $H$ can be written, up to normalization, as $\rho = e^{-\beta H}$ for some inverse temperature $\beta = 1/T>0$. This state can also be written as  
\beq
\rho = e^{-\beta H/2} ~Â {\mathbb I} ~Â e^{-\beta H/2},
\eeq
where ${\mathbb I}$ is the identity operator which, for a lattice system, decomposes as the tensor product of identities for every site. As such, this state can be understood, at least intuitively, as an imaginary time evolution with respect to both the 
vector (``ket'') and the dual vector (``bra'') degrees of freedom. The identity ${\mathbb I}$ can easily be written as a \emph{projected entangled pair operator (PEPO)} of unit bond dimension. Moreover, ${\mathbb I}$ is also the mixed quantum state (again up to normalization) at $\beta=0 ~ (T = \infty)$. Thus, in order to obtain a thermal state at finite temperature $T$, one can simply cool down, i.e. anneal, this state. Ideally, this is done as ``slowly'' 
as possible in the sense of a large number of steps  in order to avoid getting stuck in metastable states. To simulate this procedure, one can divide 
the final anticipated $\beta$ into positive integer $m$ ``temperature slices" $\Delta \beta \ll 1$, such that $m  \Delta \beta = \beta$. The mixed state can then be written as  
\beq
 \rho = (e^{- \Delta \beta H})^{m/2} ~ {\mathbb I} ~ (e^{- \Delta \beta H})^{m/2},  
\label{eq2}
\eeq
see Fig.~\ref{pepo1}(a). {The exponential $e^{- \Delta \beta H}$ can be well approximated 
via a Suzuki-Trotter expansion. For instance, for a local Hamiltonian made of 
two-body non-commuting terms $H = \sum_{i,j} h_{i,j}$ with $\|h_{i,j}\|=O(1)$,
one has 
{$\| e^{- \Delta \beta H} - \prod_{i,j} e^{-\Delta \beta h_{i,j}} \|= O(n^2 \Delta \beta^2)$ for a system size $n$.}}
In order to simulate Eq.~\eqref{eq2}, we build on the vectorization approach from 
Ref.~\cite{Kshetrimayum}, i.e., we consider 
vectors and dual vectors
 together, and implement the thermal state as an imaginary-time evolution of a vectorized mixed state $\ket{\cdot}_\sharp$ as  
 \beq
 \ket{\rho}_\sharp = (e^{- \Delta \beta H \otimes \mathbb{1}})^{m/2} (e^{- \Delta \beta \mathbb{1} \otimes H^T})^{m/2}  \ket{{\mathbb I}}_\sharp,
 \eeq
 using an isomorphism between mixed states $\rho$ and state vectors $\ket{\rho}_\sharp$. 
This equation means that the calculation of the thermal state is formally equivalent to the imaginary-time evolution of a state vector
$\ket{{\mathbb I}}_\sharp$. The key point is that, for a Hamiltonian $H$ consisting of local interactions on a two-dimensional lattice, we can actually implement such an evolution using the full machinery of algorithms that has been developed for pure states. We do this by choosing to work with the so-called ``simple update" \cite{simpleupdatejiang} scheme for tensor updates with  $\Delta \beta = 10^{-4}$, and CTM techniques \cite{ctmroman2009,ctmroman2012} for the evaluation of expectation values. The simple update scheme assumes a `mean-field' like environment while making the tensor updates. The effect of the whole environment is then included while calculating the observables using the CTM techniques as illustrated in Fig. \ref{pepo1}(c).

Our method has a number of advantages with respect to other approaches in two spatial dimensions. First, and as said above, we can straightforwardly apply everything we already know about imaginary-time evolution of PEPS, including different schemes of tensor updates, this one turning out to be particularly feasible. Second, it is much more efficient than schemes based on TNs for   purifications of the mixed state \cite{PhysRevB.92.035152,PositiveMPO,PhysRevLett.93.207204}, 
because we only need a single 2D layer of tensors to describe expectation values and effective environments, as compared to the double-layer approach of purification schemes. The scaling in our case is $O(dD^4+\chi^2D^4+\chi^3D^3)$ where $d$ and $D$ corresponds to the physical and bond dimension of the PEPO and $\chi$ is the bond dimension of the CTM. Moreover, it is known that such purifications may lead to lower accuracies (on top of slower performance) in TN algorithms because of intrinsic limitations \cite{Gemma,UndecidableMPO}. 
Additionally, we choose to use the ``simple update" of tensors throughout the evolution because it is particularly fast and efficient when dealing with gapped systems. The simple update has a cost of $O(d^4D^5+d^{12}D^3)$. Given the large degree of complexity of simulating two-dimensional thermal states, we find that this degree of efficiency is  important, and it makes the algorithm way faster than those developed using other types of tensor updates, even without purifications \cite{shiju}. All in all, and as we shall see, our method provides fast and accurate results for the studied systems not just in the limiting cases of infinite and zero temperature, but also in the intermediate-temperature regime, where strong thermal and quantum fluctuations are simultaneously present. 

\begin{figure}
	\includegraphics[width=0.36\textwidth]{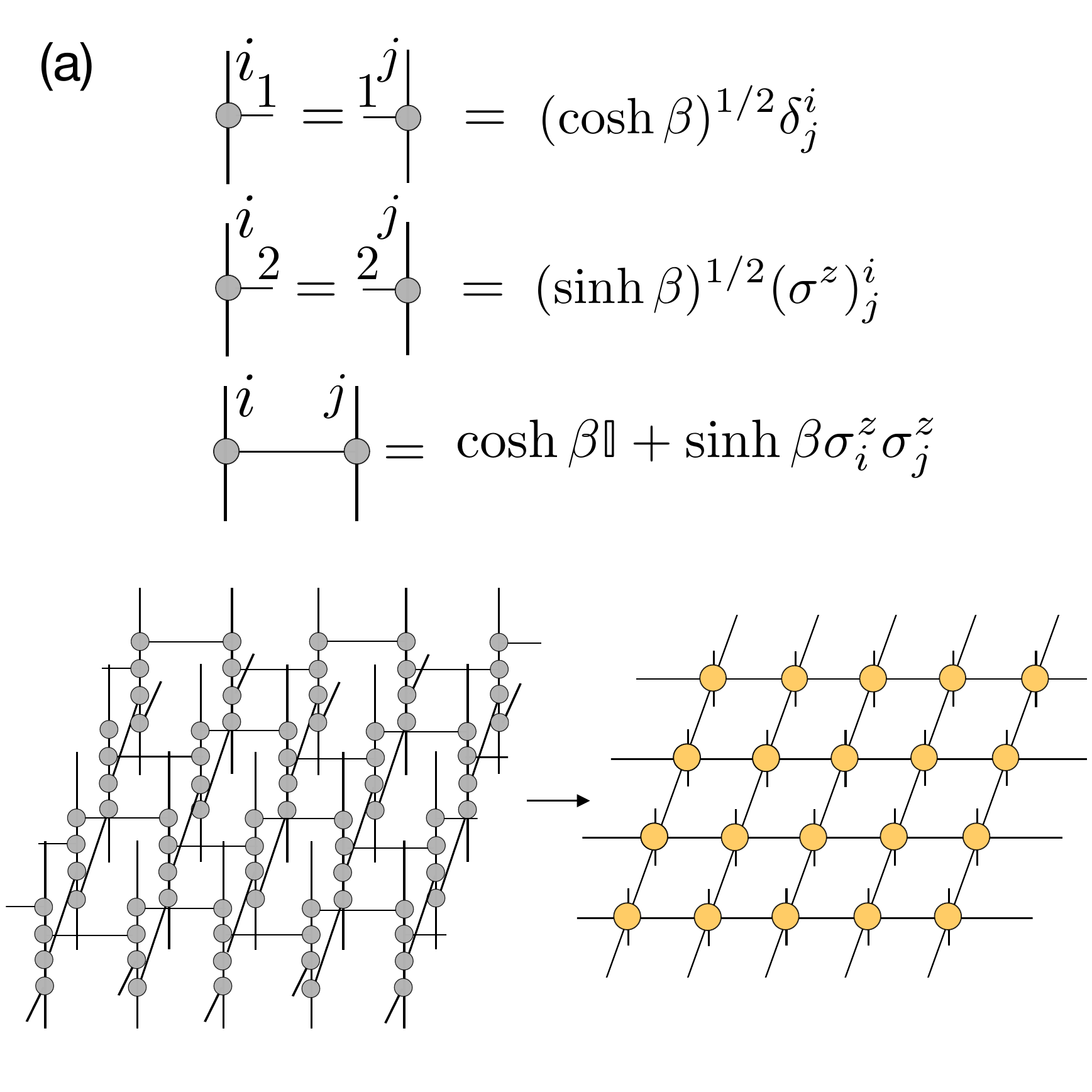}
    \includegraphics[width=0.36\textwidth]{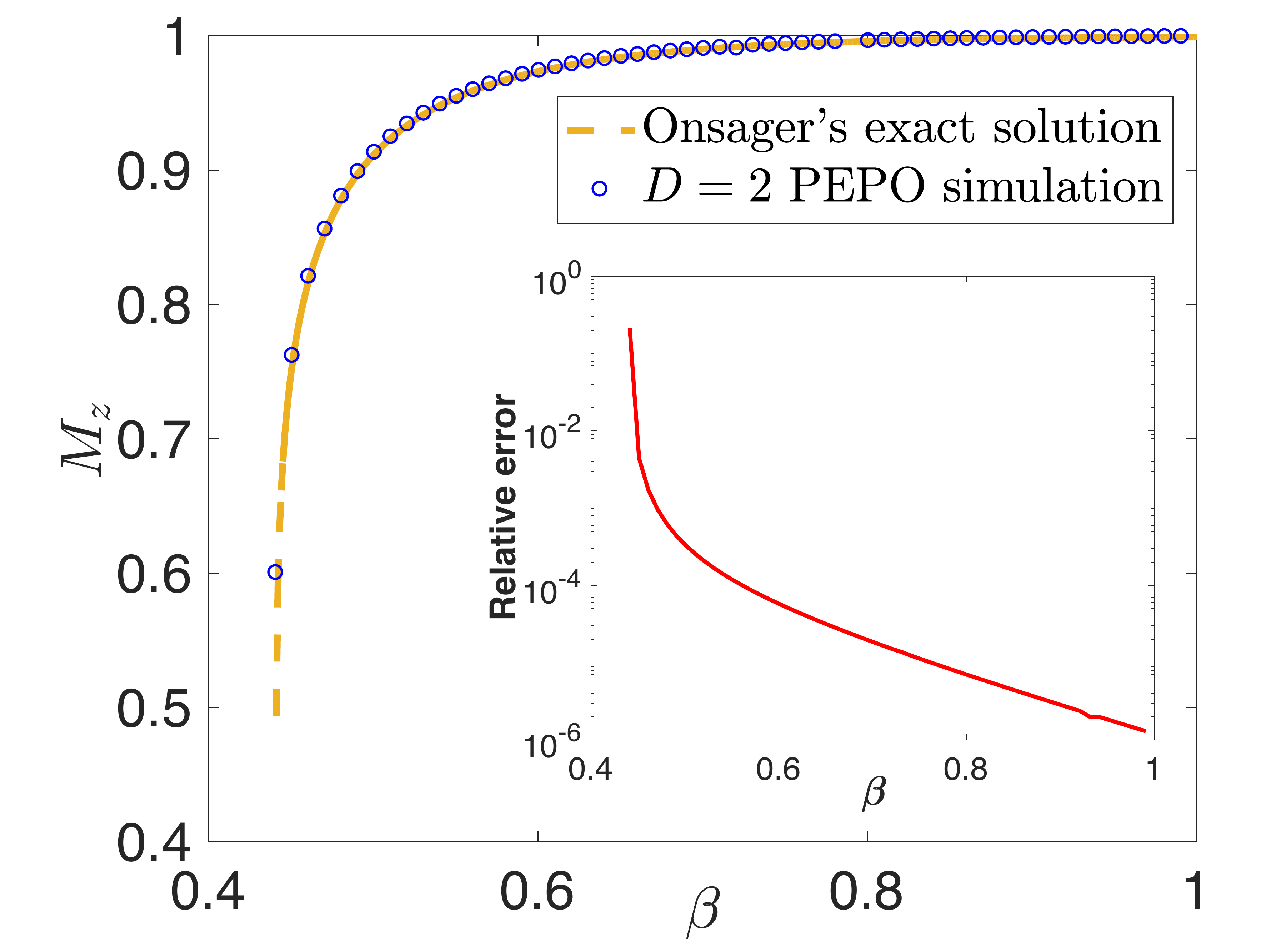}
	\caption{Finite-temperature phase diagram of the ferromagnetic Ising model on an infinite square lattice. The red curve is obtained by using our approach with bond dimension $D=2$. The blue curve is the exact solution obtained by Onsager \cite{Onsager44}. In the inset we show the relative error. 
	Notice that the error increases around the critical value of $\beta$, as expected from CTM contraction methods when dealing with large amount of correlations. }
	\label{isingpepo}
\end{figure}

\emph{Results for the Ising model.} We start by benchmarking the validity of our approach with the \emph{ferromagnetic Ising model} on an infinite square lattice at finite temperature, which can be exactly solved as famously proven by Onsager \cite{onsager}. It is easy to show that the thermal density matrix of this model can be written as a PEPO with bond dimension $D = 2$. In order to see this, consider the Hamiltonian of the model, given by 
\beq
H  =-\sum_{\langle i, j \rangle}  \sigma_i^z \sigma_j^z
\eeq
where $\sigma^z$ is the Pauli-$Z$ matrix supported on site $j$. The thermal density matrix at inverse temperature $\beta$ can be written as
\beq
\begin{split}
\rho &
= e^{\beta \sum_{\langle i,j \rangle}\sigma_i^z \sigma_j^z} = \prod_{\langle i, j \rangle}(\cosh \beta ~ {\mathbb I} + \sinh \beta \sigma_i^z \sigma_j^z ).
\end{split}
\label{Tisingpepoeq}
\eeq
The expression in Eq.~\eqref{Tisingpepoeq} is nothing but a product over the links of two-site 
\emph{matrix product operators (MPO)} with bond dimension two. The product of all such 
MPOs on a square lattice amounts to an exact PEPO with bond dimension $D=2$, as claimed. Thus, the fact that this model allows for an exact PEPO representation for any $\beta>0$ implies that it is an excellent model to benchmark our numerical method. Following this idea, we obtain the PEPO for the thermal density matrix by using our numerical technique. The results are depicted in Fig.~\ref{isingpepo}, where one can see that our algorithm produces remarkably good results when compared to the exact solution. 

\emph{Results for Bose-Hubbard models.} 
Next, we apply our method to study the finite-temperature properties of a non-integrable model, specifically the Bose-Hubbard model on the infinite square lattice. The model is itself very relevant in the context of ultra-cold atom experiments with optical lattices. Moreover, real-life laboratory conditions imply small thermal fluctuations, which we can conveniently target via our method. The ground state properties of this model have been widely studied with a variety of methods. 
For finite-temperature, quantum Monte Carlo methods are still applicable providing benchmark results \cite{0710.2703,PhysRevA.70.053615}, 
but not in case of a sign-problem, in contrast to
the method used here
\cite{CorbozPEPSFermions}.

\begin{figure}
	\includegraphics[width=0.38\textwidth]{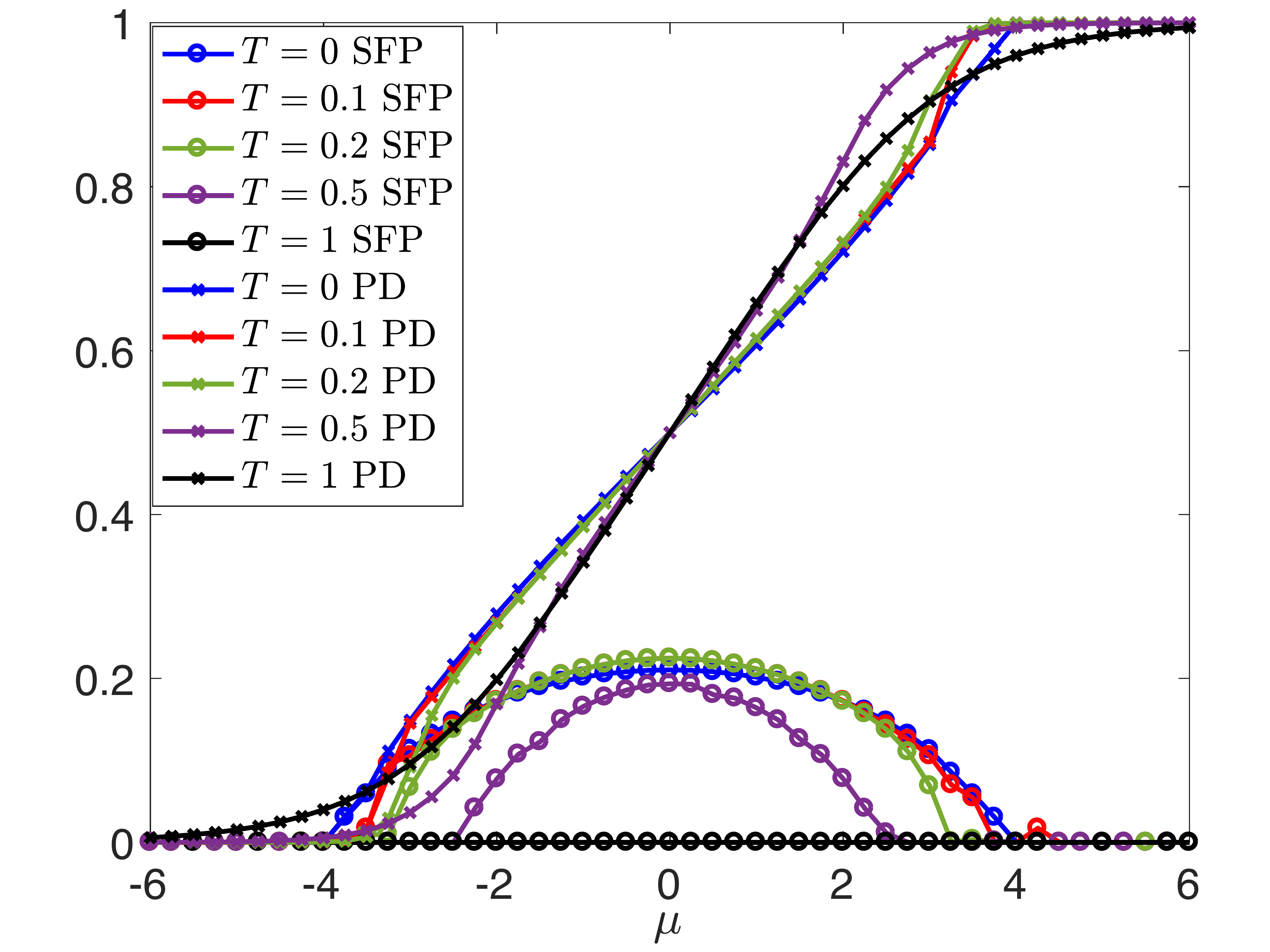}
    \includegraphics[width=0.38\textwidth]{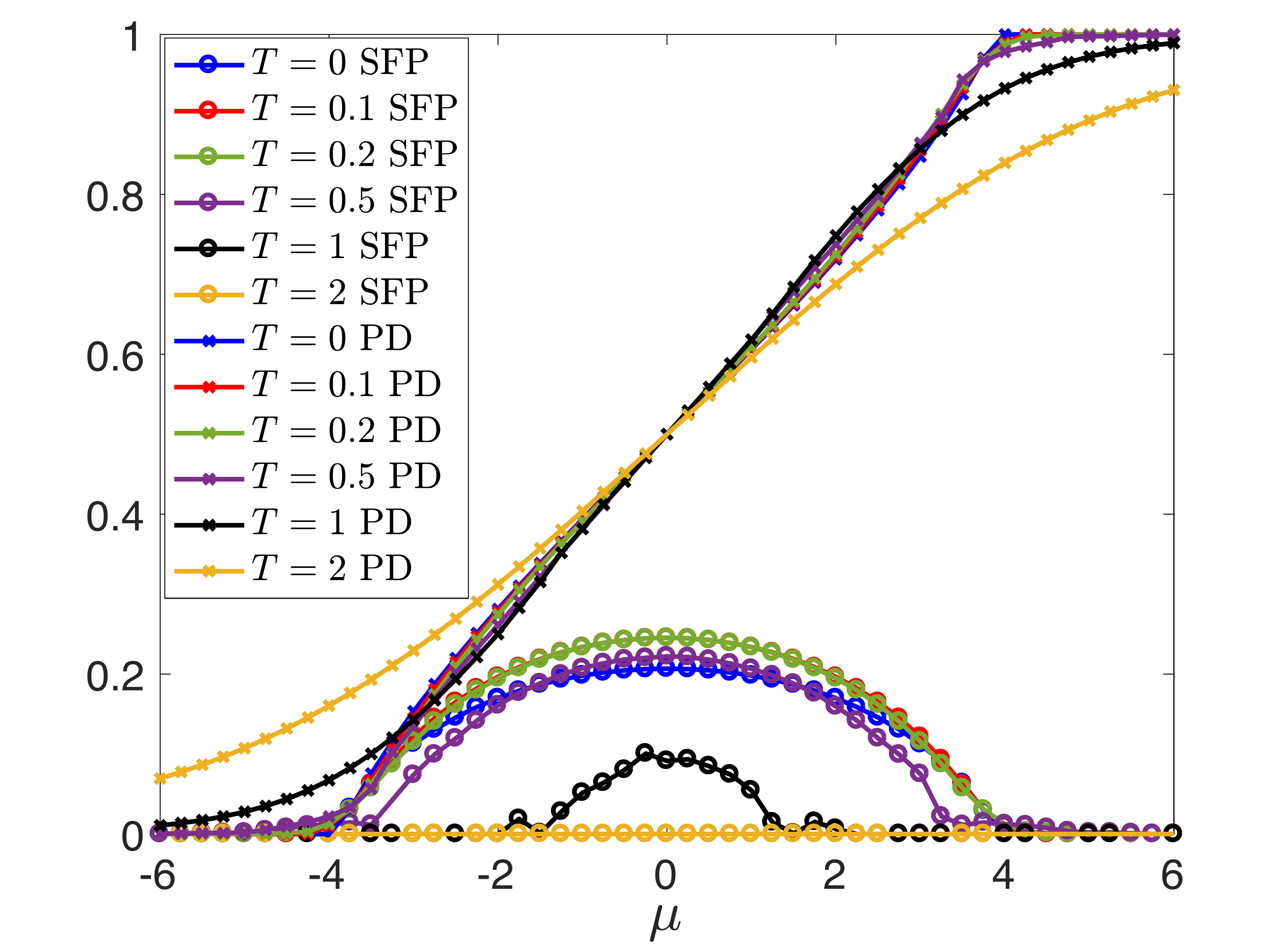}
	\caption{Phase diagram of the hard-core Bose Hubbard model for different values of $T$. The $x$-axis corresponds to $\mu/J$ with $J=1$ where $\mu$ is the chemical potential and $J$ is the hopping parameter. Results are for bond dimensions $D=2$ (above) and $D=3$ (below). Circles correspond to the superfluid parameter (SFP) and crosses to the number of particles (PD) per site. We can see that the superfluid phase shrinks in size as we increase the temperature until it disappears completely at $T\approx 1$ and $T \approx 2$ for $D=2$ and $D=3$ simulations respectively.}
	\label{HCBHD2}
\end{figure}

We start by considering the hard-core limit, i.e., the situation when only one boson per site is allowed at most and therefore we have a local Fock space of local dimension two (empty/occupied state). 
The Hamiltonian is given by  
\beq
H = - J \sum_{\langle i, j \rangle}(a_i^{\dagger}a_j + a_j^{\dagger}a_i) - \sum_i\mu\hat{n}_i
\eeq
where the real $J$ is the hopping strength, $a_i^{\dagger}$, $a_i$ are the hard-core bosonic creation and annihilation operators at site $i$, $\mu$ is the chemical potential and $\hat{n}_i :=  a_i^{\dagger}a_i$ is the number density operator. In this limit the model is equivalent to the quantum spin-1/2 $XY$ model in the presence of an external field. This is because the hard-core bosonic creation and annihilation operators can be written in terms of spin raising and lowering operators as $a_i^{\dagger} := \hat{S}_+ = \frac{1}{2}(\sigma_i^x + i\sigma_i^y)$,  $a_i := \hat{S}_- = \frac{1}{2}(\sigma_i^x - i\sigma_i^y)$ and $\hat{n}_i := a_i^{\dagger} a_i = \frac{1}{2}(\sigma_i^z + \mathbb{I})$. Using this mapping, the model can be written in terms of the spin-1/2 Hamiltonian 
\beq
H = -\frac{J}{2}\sum_{\langle i, j \rangle} \left(\sigma_i^x \sigma_j^x + \sigma_i^y \sigma_j^y \right) + \frac{\mu}{2} \sum_i \sigma_i^z.
\eeq
We have computed the thermal phase diagram of this model using our annealing algorithm, and for completeness and comparison, also the zero-temperature properties with \emph{infinite PEPS (iPEPS)} \cite{simpleupdatejordan,iPEPSBH}. The particle density $\langle \hat{n}_j \rangle$ is shown in Fig.~\ref{HCBHD2} for bond dimension $D = 2$ and different temperatures $T$, as a function of the chemical potential $\mu$ and taking $J = 1$. For very low temperatures $T \approx 0$, we see that $\langle \hat{n}_j \rangle = 0$ for $\mu \leq -4$ and $\langle \hat{n}_j \rangle = 1$ for $\mu \geq 4$. These regions in the phase diagram contain thus an integer number of bosons per site and correspond to 
a \emph{Mott insulating (MI)} phase, whereas the region $-4<\mu < 4$ is in the \emph{super-fluid (SF)} phase. 
As the temperature increases, the size of the SF phase shrinks, until eventually disappearing around $T \approx 1$. We conclude that the SF-MI transition is not just a property of the zero-temperature state, but it also shows 
clear signatures at low but finite temperatures. At larger temperatures $T > 1$, though, we see that this transition is no longer  visible. Additionally, we have 
computed the order parameter $\rho_0 = |\langle a_j \rangle|^2$, which reveals the broken $U(1)$ symmetry of the SF phase and provides information on the ``condensate fraction". This is also shown in Fig.~\ref{HCBHD2}. Clearly, the order parameter is vanishing for large enough temperatures, as expected for the breaking down of the condensate when the temperature of the system increases. The finite bond dimension $D$ seems to induce a finite correlation length in the system, thus imposing an effective length scale. The physics that we are observing, thus, may also be more compatible in some situations with the expected behavior for a finite system: this concerns, e.g., the appearance of a finite condensate fraction at $T\neq 0$ \cite{Carrasquilla_Rigol2012}. The correct thermodynamic limit is then recovered doing finite-$D$ scaling.

\begin{figure}
    \includegraphics[width=0.38\textwidth]{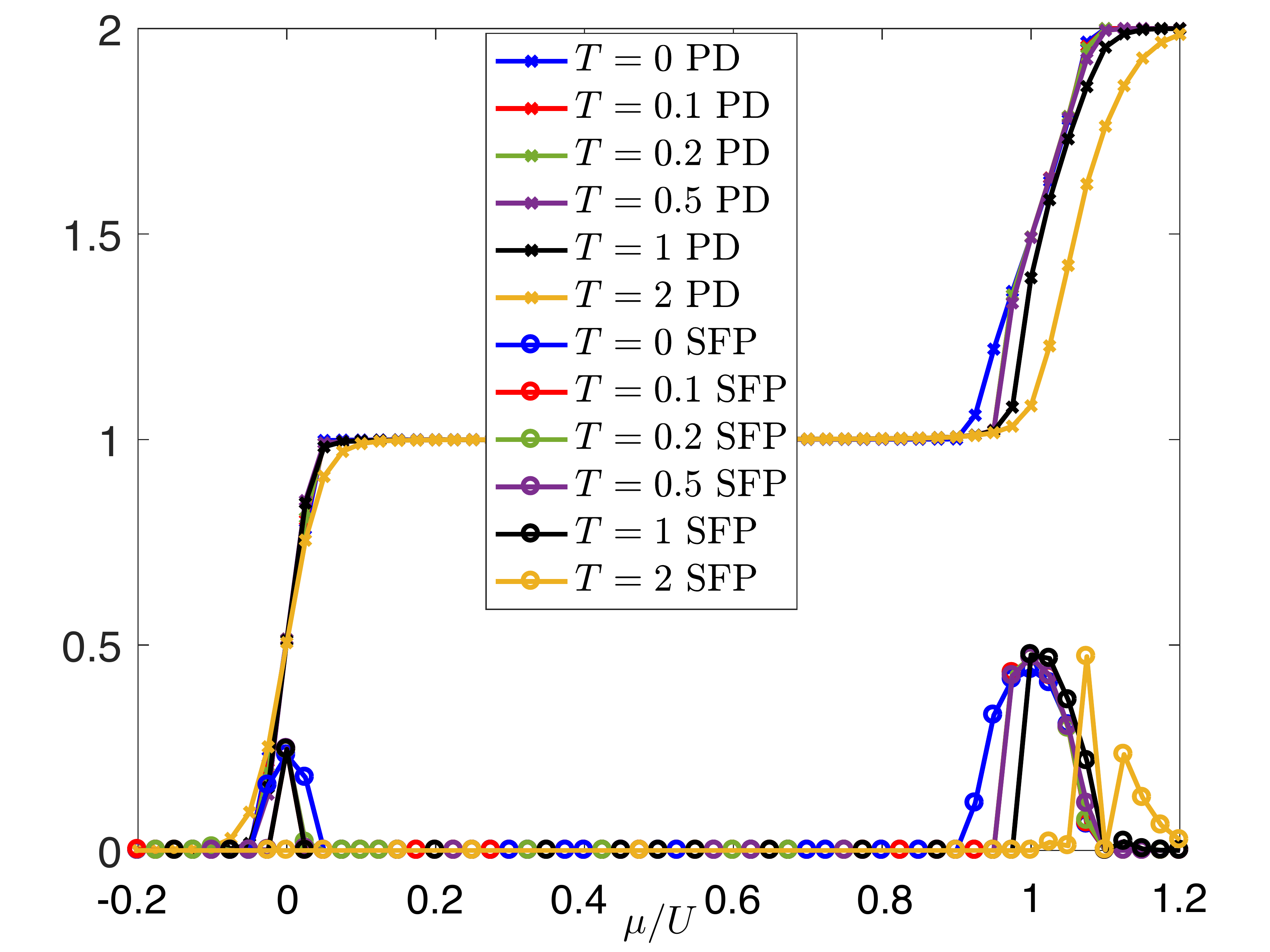}
    \includegraphics[width=0.38\textwidth]{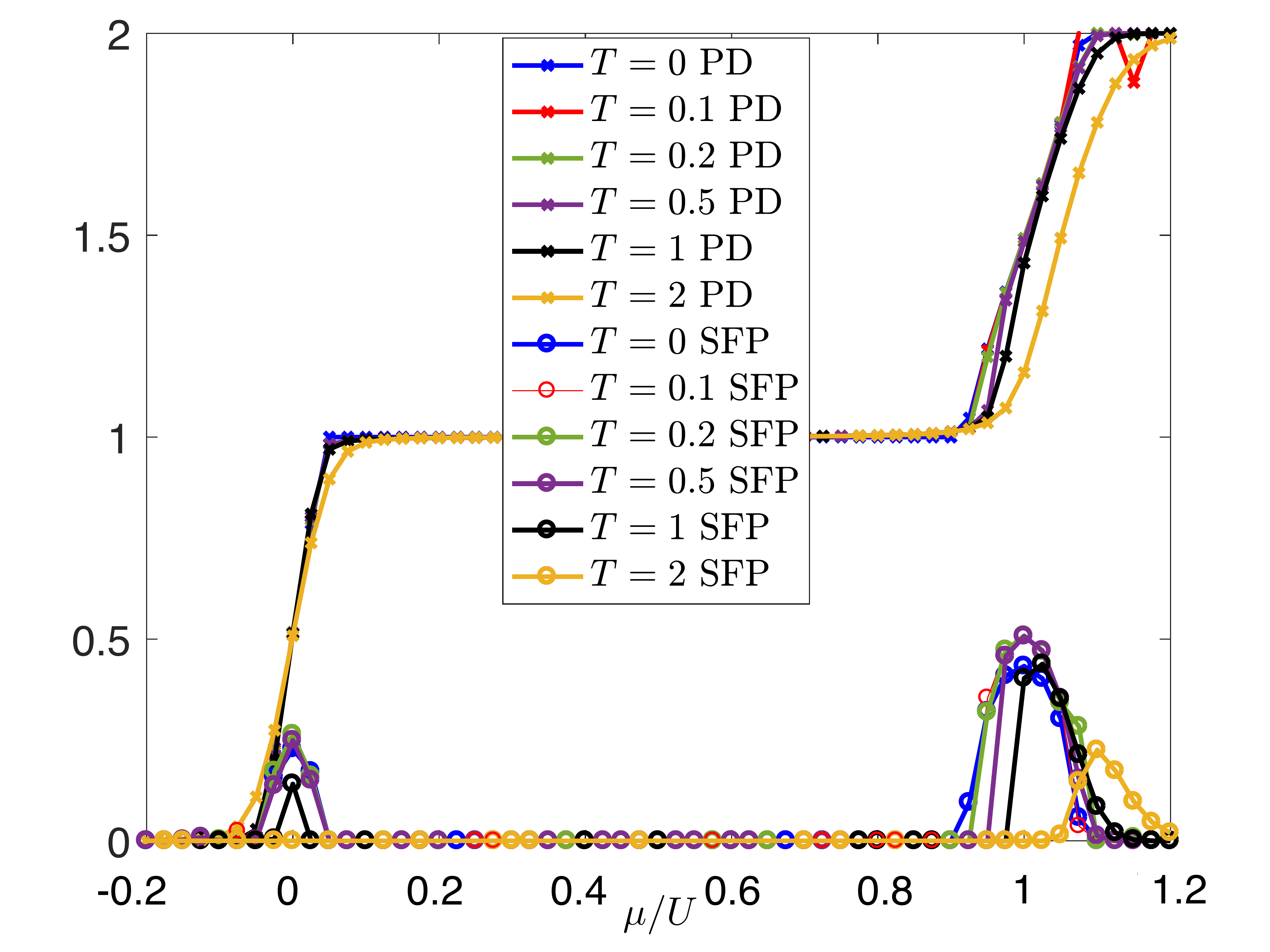}
	\caption{Phase diagram of the three-level Bose Hubbard model, a softer version of the hard core constraint for different values of $T$, for $U/J = 100$. The plots with circular data marks correspond to the superfluid parameter which is the square of the expectation value of $a_i$. The plots with cross data marks correspond to the number of particles per site. The Mott insulating phase has integer particle density ($n=1$ and $n=2$) while the superfluid phase (appearing in between/between plateaus) breaks particle number conservation symmetry. At sufficiently high temperatures, the superfluid disappears in both intermediate regions, though most slowly in the second one. The upper panel is for $D=2$ and the lower panel for $D=3$. This time we observe very similar features for both values of the bond dimension.}
	\label{SCBH}
\end{figure}

We now turn to relaxing the hard-core condition in order to allow for larger particle numbers per site. Here, we consider the simplest soft-core example, allowing up to two bosons per site, i.e., local Fock space of dimension three. 
This could interestingly be achieved as the limiting case of strong three-body dissipation~\cite{Daley2009,Roncaglia2010,DaleyOpenReview} or by a proper tuning of the super-exchange regime of spin-1 particles~\cite{Mazza2010}.
Such a restricted Bose-Hubbard model might disclose paths towards interesting fractional quantum Hall states, once decorated with synthetic magnetic fluxes.
The Hamiltonian is given by 
\beq
H = -J \sum_{\langle i,j \rangle}(a_i^{\dagger}a_j + a_j^{\dagger}a_i) - \mu \sum_i \hat{n}_i + \frac{U}{2} \sum_i \hat{n}_i(\hat{n}_i - 1), 
\eeq
which takes the same form as  the hard-core Hamiltonian, but with the extra term of the on-site repulsive density-density interactions of strength $U$ (and which are zero in the hard-core limit $U \rightarrow \infty$). We perform a similar study as the one done in the hard-core case, focusing on the small-hopping regime  for demonstrative purposes. In particular, we set the parameters to $J=1$, $U=100$, and we study the phase diagram as a function of the chemical potential $\mu$. Our results are shown in Fig.~\ref{SCBH}, where we find similar results as for the hard-core limit, but with MI phases at $\langle \hat{n}_j \rangle = 0, 1, 2$ and SF phases at intermediate regions between these occupation numbers which shrink in size as the temperature increases and eventually tend to disappear at sufficiently large $T$.
Interestingly, the fluctuations of the on-site occupation number, $\langle {\hat n}_j^2\rangle-\langle  {\hat n}_j\rangle^2$ (not shown here), retain their qualitative behavior at all considered temperatures:
i) They vanish where $\langle {\hat n}_j\rangle$ exhibit plateaus, thus signaling a certain robustness of the Mott features versus thermal fluctuations.
ii) They are sizeable in the intermediate regions, even when the order parameter $\rho_0$ vanishes, thus indicating that the system is compressible but not coherent there.
For example, we have found that for $\mu/U=0.4$, the variance is of the order of $10^{-3}$ confirming the robustness of the MI phase even at $T=2$.
Moreover, the results of Fig.~\ref{SCBH} also justify our approximation of low occupation number (i.e., up to two bosons per site, so that for larger $\mu$ the simulations are no longer reliable), for the studied regime of chemical potential \footnote{It is interesting to note a shift in the right-hand-side SF region at $T = 2$; the precise
origin of this and whether this can be derived from the choice of truncation will be explored in detail in future work.}.


\emph{Conclusions.}  In this work we have introduced and discussed an efficient tensor network algorithm to compute finite-temperature properties of two-dimensional quantum lattice systems in the thermodynamic limit. The method uses an annealing procedure which is simulated via vectorization of projected entangled pair operators describing the thermal state. We have benchmarked the algorithm with the exact solution of the 2D Ising model on the square lattice, and applied it subsequently to study finite-temperature properties of the Bose Hubbard model on the square lattice, in the hard-core limit as well allowing up to two bosons per site. Our method is fast and easy to implement,
whenever PEPS expertise is available for  ground state  calculations. Because of this, it is our belief that it will become a versatile tool in future studies of finite-temperature properties of two-dimensional quantum matter, as well as in the benchmarking of optical-lattice experiments in two dimensions under real-life laboratory conditions. 

\emph{Acknowledgements.} We thank the ERC (TAQ), the DFG (CRC 183 projects B1 and B2, 
 EI 519/7-1, and EI 519/14-1),
and the Templeton Foundation for support. 
This work has also received funding from the European Union's Horizon 2020
research and innovation programme under grant agreement No 817482 (PASQUANS).
Discussions with Juan Bermejo-Vega and Alexander Nietner are also acknowledged. Part of the simulations were done at the MOGON cluster at JGU.

\bibliographystyle{apsrev4-1}

%

\end{document}